\documentclass[english,aps,prb,reprint,showpacs,a4paper,preprintnumbers]{revtex4-1}
\usepackage{ae,aecompl}
\usepackage[T1]{fontenc}
\usepackage[latin9]{inputenc}
\usepackage{graphicx}
\usepackage{natbib}

\makeatletter

\providecommand{\tabularnewline}{\\}

\@ifundefined{showcaptionsetup}{}{%
 \PassOptionsToPackage{caption=false}{subfig}}
\usepackage{subfig}
\makeatother

\usepackage{babel}
\usepackage{fancyhdr}
\begin{document}

\title{A permutation invariant collective variable to track and drive vacancy
dynamics in simulations of solids}

\author{Jan M. Knaup}

\email{jan.knaup@bccms.uni-bremen.de}

\author{Michael Wehlau}

\author{Thomas Frauenheim}
\email{frauenheim@bccms.uni-bremen.de}

\affiliation{Bremen Center for Computational Materials Science, University of
Bremen, Am Fallturm 1, D-28359 Bremen, Germany}
\begin{abstract}
Vacancy dynamics in oxides are vital for understanding redox reactions
and resulting memristive effects or catalytic activity. We present
a method to track and drive vacancies which we apply to metadynamics
simulation of oxygen vacancies (V$_{\mathrm{O}}^{2+}$) in rutile, demonstrating
its effectiveness. Using the density functional based tight binding
method, it is possible to explore the free energy hyperplane of oxygen
vacancies in TiO$_{2}$. We show that the migration of V$_{\mathrm{O}}^{2+}$
in TiO$_{2}$ is governed by the jump with the highest degree of topological
interconnection. Free energy profiles are consistent with minimum energy
paths.
\end{abstract}


\preprint{cite as Phys. Rev. B. {\bf 88}, 220101(R) (2013)}

\maketitle

Transition metal oxides exhibit a highly complex behavior, with strong
coupling between electronic and ionic transport processes, making them
highly interesting
as electronic materials beyond their traditional use as dielectrics. 
The most prominent example of such effects is
the memristive behavior\cite{Kwon-nn-5-148-2010,*Park-IEDL-32-197-2011,*Strukov-N-453-80-2008,*0957-4484-20-48-485701}
of TiO$_{2-x}$ and many other substoichiometric oxides\cite{Waser-AM-21-2632-2009}.
In titania, the memristive behavior is linked to the formation and
dissolution of conducing filaments\cite{Kwon-nn-5-148-2010,*Park-IEDL-32-197-2011,*Strukov-N-453-80-2008,*0957-4484-20-48-485701}
or layers\cite{pickett:074508} of room-temperature metallic Ti$_{n}$O$_{2n-1}$
Magn\'eli phases\cite{Bursill-PSSC-7-177-1972,*Marezio1973213,*PhysRevB.77.104104,*PhysRevB.79.245133}
in the insulating TiO$_{2}$. These phases of titania are modifications
of the rutile crystal structure in which oxygen vacancies (V$_{\mathrm{O}}$)
are ordered in slip-planes. In order to understand their formation
and dissolution it is essential to understand the dynamics of oxygen
vacancies aggregating and dispersing.

Yet, the memristive effect is far from being the only interesting
physics of metal oxides in which vacancy dynamics play a crucial role.
Catalytic effects at oxide surfaces can rely on the presence and ability
to replenish surface vacancies\cite{Campbell29072005,Bikondoa2006/,*doi:10.1080/01614949208021918,*Wanbayor201224,*wanbayor:104701,*PhysRevLett.109.136103}
or O ion transport through the catalyst\cite{Dow1996155,Campbell29072005}.
To be efficient, the latter relies on vacancies and can in fact be
mapped to a vacancy transport process in the reverse direction.
Oxide based (photo)catalysis is regarded as one of the most promising
fields of renewable energy conversion.

These examples show that complex physical processes at the surface
and in the bulk of reducible metal oxides require intimate understanding
of the dynamical behavior of V$_{\mathrm{O}}$ defects and the relevant
driving forces. Many of these processes are rare events, i.e. their
activation energy is larger than the average thermal energy at relevant
conditions, which makes them inaccessible to unbiased molecular dynamics
simulation. Techniques for rare event simulation\cite{Laio01102002,PhysRevLett.109.150601,PhysRevLett.109.020601,PhysRevB.57.R13985}
rely on reaction coordinates describing the transition of interest.
In most cases, the reaction coordinate is required to be continuous
and at least piece-wise differentiable with respect to the atomic
positions $\vec{r_{i}^{\mathrm{lat}}}$. Achieving this for a vacancy
coordinate $\vec{r_{\mathrm{v}}}$ is difficult, as vacancies are
not directly simulated objects but an emergent property. In the strictest
sense, the position of a vacancy cannot be continuous, as it is defined
as an unoccupied lattice position. Yet, since a vacancy can change
its position, albeit indirectly, a generalization of the vacancy position
to allow the description of the transition is possible. So far, studies
of vacancy migration focus on single hopping events, tracing the position
of the moving neighbor atom, and implicitly defining the vacancy motion
as the opposite of the ion motion\cite{PhysRevB.75.073203,Asaduzzaman2010,KnaupMRS2012,B502507A}.
This auxiliary definition suffers from the conceptual problem, that
at or at least close to the transition state the vacancy defined as
``where no ion is'' coincides with the position of the moving ion.
Of greater practical impact is that focusing on a single ion when
driving the vacancy severely limits the search space for the vacancy
transition path. The \emph{a-priori} assumption that the motion of
only the selected atom alone leads to the vacancy motion precludes
any collective effects.

One approach to capture the emergent nature of vacancies is to analyze
the simulated volume for voids and designate such spaces as vacancies
based on their size (e.g. \cite{prb-76-235201-2007,PhysRevB.84.094124}).
While highly flexible and useful to detect an unknown number of vacancies
in a matrix of unspecified crystal structure, the approach is computationally
cumbersome and can neither yield a continuous $\partial\vec{r_{\mathrm{v}}}/\partial\vec{r_{i}^{\mathrm{lat}}}$
at acceptable computational effort, nor can it describe the crucial
hopping events.

We return our focus to the discrete reference positions. In the transition
state, the moving ion is equally far from two lattice positions%
\footnote{For the sake of simplicity and without loss of generality we assume
for a moment that the transition barrier is symmetrical between the
adjacent minima.%
}, which suggests having two vacancies while only one ion is missing.
This leads us to the desired generalization of the vacancy concept
allowing for the continuous tracking of vacancy motion: Instead of
a Boolean assignment of ions to lattice positions, ions may impart
a certain amount of \emph{occupiedness} to lattice sites, ranging
from $0\ldots1$ depending on their distance. In this picture the
ion at the transition state partially occupies both adjacent lattice
positions. $\vec{r_{\mathrm{v}}}$ is then defined as the average
of the lattice positions weighted by unity minus their \emph{occupiedness}.
In contrast to coordinates derived from specific ion positions, this
concept is inherently invariant under permutation of the ionic labels,
removing many \emph{a-priori} constraints. We therefore call this
coordinate \emph{Permutation Invariant Vacancy Location Trajectory},
in short \emph{PIVoT}. Mathematically, the position of the vacancy
is determined from the externally supplied reference lattice positions
$\vec{r_{i}^{\mathrm{lat}}}$ and the ionic positions $\vec{r_{j}^{\mathrm{ion}}}$
by a weighted average
\begin{eqnarray}
\vec{r_{\mathrm{v}}}= & \sum_{i} & \vec{r_{i}^{\mathrm{lat}}}\frac{w_{i}}{\sum_{i}w_{i}}\left[\frac{G_{i}}{\sum_{i}G_{i}}\right]\label{eq:pivot}\\
w_{i}= & 1 & -\frac{1}{\exp\left(\left(r_{\mathrm{min},i}-d_{0}\right)/\tau\right)+1}\\
G_{i}= & \exp & \left(\frac{-\left|\vec{r_{i}^{\mathrm{lat}}}-\vec{r_{\mathrm{v}}}\right|^{2}}{2\left(\sigma_{\mathrm{G}}\right)^{2}}\right)\label{eq:Gaussian}\\
r_{\mathrm{min},i}= & \min_{j} & \left(\left|\vec{r_{i}^{\mathrm{lat}}}-\vec{r_{j}^{\mathrm{ion}}}\right|\right)
\end{eqnarray}
where the weighting function $w_{i}$ is a mirrored Fermi-Dirac distribution
function with the switchover distance $d_{0}$ and a steepness parameter
$\tau$. It was chosen for being able to control the smooth transition
with convenient parameters. For tracing a vacancy in the post-processing
of existing trajectories, $d_{0}$ and $\tau$ are chosen such that
the weight of a lattice position remains almost zero as long as the
closest ion does not stray outside the range of thermal oscillation.
For use as a collective variable (CV) in meta dynamics, the derivative of the CV with respect
to atomic coordinates would be zero close to equilibrium and the algorithm
could not drive the dynamics. In this case, $d_{0}$ and $\tau$ are
chosen to yield a smoother transition and weight thermal displacements.
Since the global averaging of all $\vec{r_{i}^{\mathrm{lat}}}$ will
now draw $\vec{r_{\mathrm{v}}}$ towards the center of gravity of
the $\vec{r_{i}^{\mathrm{lat}}}$, the term $G_{i}$ needs to be applied
to additionally weight the $\vec{r_{i}^{\mathrm{lat}}}$ by their
distance from $\vec{r_{\mathrm{v}}}$. In this case Eq.~\ref{eq:pivot}
has to be solved self consistently, with the $\sigma$ parameter chosen
so that the FWHM of $G_{i}$ falls between the 2nd and 3rd shells
of the moving species.

In our simulations of V$_{\mathrm{O}}^{2+}$ diffusion in TiO$_{2}$we
calculate energies and forces using the self-consistent charge version
of the density-functional based tight-binding method~\cite{ThFRAUENHEIM-GS-ME-ZH-GJ-DP-SS-RS-PSSb-217-41,*JPCA-111-5678-2007}
(SCC-DFTB) with the tiorg parametrization~\cite{doi:10.1021/ct900422c}.
We iterate SCC until charge errors are $<10^{-8}$ e$^{-}$. Supercells
between 2x2x3 and 4x4x6 rutile units, each containing a single V$^{2+}{}_{\mathrm{O}}$
defect were simulated, using a 2x2x2 Monkhorst-Pack $k$-point set\cite{PhysRevB.13.5188}.
The 2+ charge state was chosen based on the results in ref.~\onlinecite{PhysRevB.86.195206}.
We compensate for the defect charge by applying a uniform background charge.
Meta dynamics\cite{Laio01102002} simulations were carried out in
the NVT ensemble at 600~K ensured by a Nose-Hoover thermostat\cite{Nose1984,PhysRevA.31.1695}
for the ions, and using the PIVoT collective variable projected onto
the (001) lattice direction. We choose $d_{0}=1.5$~\AA\  and $\tau=0.4$~\AA\ 
to ensure reasonable derivatives.

To understand the behavior of the PIVoT coordinate in detail, we compare
the resulting vacancy trajectory to the center of gravity of the cavity
in the O sublattice resulting from the vacancy. The void is determined
by a procedure similar to cavity analysis in GeSbTe phase change materials~\cite{prb-76-235201-2007}
but explicitly without analyzing the void size. We examine volume
elements of 0.1~\AA\  edge length, marking them as empty if no O
atom is closer than 1.1~\AA, just over $\frac{1}{2}$ of the O--O
neighbor distance in rutile. During a hopping event, empty voxels
are filled by the moving O atom, while new empty elements appear behind
its original position. The center of gravity of the empty elements
transits continuously, with some thermal fluctuation. While the transition
of the vacancy can be detected this way, its time scale is underestimated
by a factor of 5, as shown by he comparison to the motion of the hopping
O atom. In contrast, PIVoT reproduces the transition time very well. 
(See Supplemental Material at [URL will be inserted by publisher] for 
a detailed comparison.)

To verify the suitability of SCC-DFTB for the simulation of oxygen
vacancy diffusion in TiO$_{2}$, we calculate the minimum energy paths
for the three symmetry-inequivalent transitions a V$^{2+}_{\mathrm{O}}$
can undergo using the nudged elastic band method for the 2x2x3 and
3x3x6 cells and adiabatic mapping for a 4x4x6 supercell. The resulting
diffusion barriers from the DFTB calculations turn out between 0.1
and 0.2~eV higher than the GGA results, as shown Table~\ref{tab:ME-Barriers},
so the agreement is reasonable. 
\begin{table}
\begin{tabular}{cccccc}
Barrier & \multicolumn{3}{c}{DFTB} & DFT\cite{PhysRevB.75.073203} & MSINDO\cite{B502507A}\tabularnewline
static & 2x2x3 & 3x3x6 & 4x4x6 & 5x4x2 & 2x3x2 (slab)\tabularnewline
\hline 
\hline 
A & 1.89 & 1.94 & 1.96 & 1.77 & -\tabularnewline
B & 1.41 & 1.03 & 0.88 & 0.69 & -\tabularnewline
C & 1.22 & 1.36 & 1.31 & 1.10 & -\tabularnewline
\hline 
free &  &  &  &  & \tabularnewline
\hline 
\hline 
A & - & - & - & - & $\sim0.96$\tabularnewline
C & 1.44 & 1.09 & - & - & $\sim1.22$\tabularnewline
\hline 
\end{tabular}

\caption{\label{tab:ME-Barriers} V$^{2+}_{\mathrm{O}}$ minimum energy diffusion
barriers and the free energy diffusion barrier of the C process in
eV. Processes A,B occur within the equatorial plane of the corresponding
Ti octahedron, process C is out-of plane. Nomenclature follows ref.~\onlinecite{PhysRevB.75.073203}.
Values for MSINDO free energy barriers are averaged between forward
and backward directions, since the (110) surface in these simulations
breaks translational invariance.}
\end{table}
 Another important result in Table~\ref{tab:ME-Barriers} is the
strong supercell size dependence of the ``B'' barrier, which is
connected to the occurrence of chains of Ti atoms linked by O double
bridges along the (001) direction in rutile\cite{KnaupMRS2012}. Careful
analysis of the geometry shows that the supercell chosen by Iddir
\emph{et.al.} for their DFT calculations is well thought out to reduce
this cell size effect. The slight increase of the A and C barriers with supercell size are
attributable to stabilization of the vacancy equilibrium configuration in larger cells.
  Recent claims of strong charge state dependence
of the V$_{\mathrm{O}}$ diffusion barriers~\cite{Asaduzzaman2010}
are dubious since the supercell size dependence is ignored there and
$k$-point sampling is inadequate.

\begin{figure*}
\subfloat[\label{fig:small-FEP-FEP} 2x2x3 FEP]{\begin{centering}
\includegraphics[width=5.5cm]{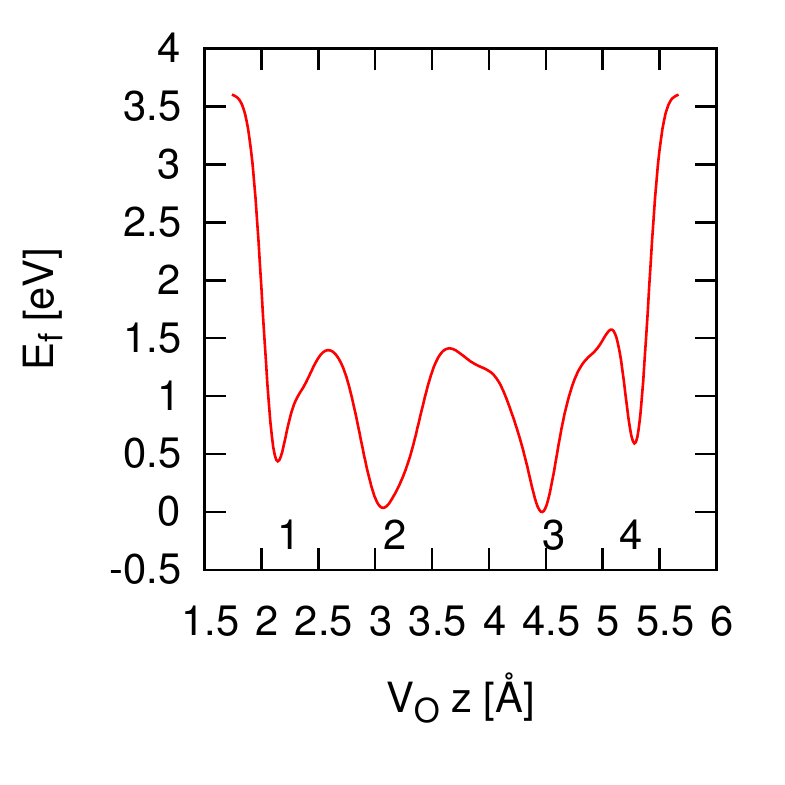}
\par\end{centering}

} \subfloat[\label{fig:small-FEP-TRJ} 2x2x3 trajectory]{\begin{centering}
\includegraphics[width=4cm]{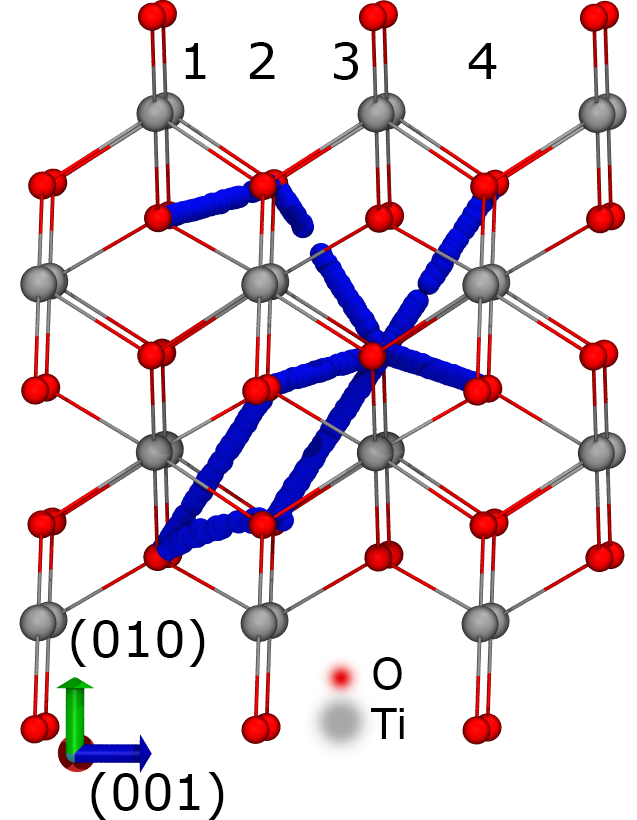}
\par\end{centering}

} \subfloat[\label{fig:big-FEP}Cutout of 3x3x6 FEP]{\begin{centering}
\includegraphics[width=5.5cm]{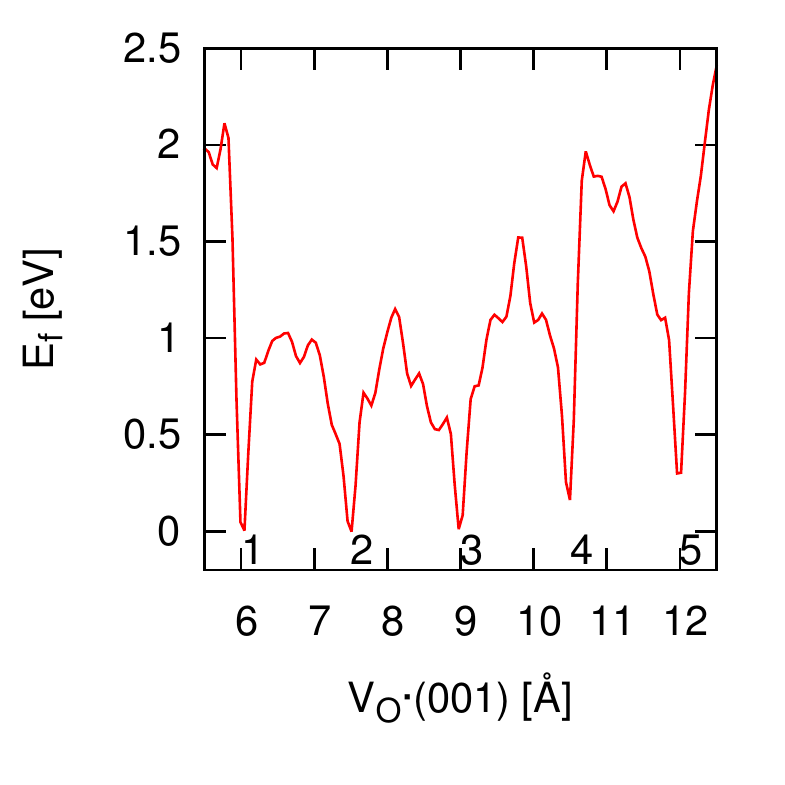}
\par\end{centering}

}\caption{\label{fig:FEPs} (color online) (a) 130,000~step free energy profile
of V$_{\mathrm{O}}^{2+}$ along (001) in 2x2x3 supercell. Wall potentials
constrain the motion of the vacancy at 1.7 and 5.8~\AA\ . (b) Vacancy
position shown every 100~fs as (dark) blue spheres, view approximately
along the (100) axis, Ti and O atoms of the minimum energy structure
of the TiO$_{2}$ cell are shown for orientation. (c) cutout of the
the FEP in the 3x3x6 supercell after 75000 steps.}
\end{figure*}
 Even though barriers from the 2x2x3 supercell are unphysically distorted,
it is edifying to compare a free energy profile obtained from this
cell to the minimum energy path (MEP). As Fig.~\ref{fig:small-FEP-TRJ} shows, the free
energy profile (FEP) in Fig.~\ref{fig:small-FEP-FEP} is averaged
over several C-jumps between adjacent TiO planes, the profiles between
minima 1--2 and 2--3 are each a thermodynamic mean of 3 distinct C-type
transitions, while the 3--4 barrier is only sampled along 2 routes.
Artificial wall potentials at $\sim1.7$ and $\sim5.8$~\AA\  were
employed to keep the vacancy from approaching the periodic boundary.
Therefore only the part between 2.5~\AA\  and 4.5~\AA\  can be
analyzed. While an A transition would be possible, it was not observed.
B Barriers will not be visible in the FEP, as the jumps are perpendicular
to the analyzed direction. The mean value of the barriers weighted
by the number of contributing transitions yields an activation free
energy of $1.44\pm0.7$~eV. This value is about 0.2~eV higher than
the MEP barrier, which is reasonable regarding the achievable accuracy
and small cell size.

Fig\@.~\ref{fig:big-FEP} shows a cutout of the FEP obtained after
75,000 meta dynamics iterations in the much larger 3x3x6 supercell.
In retrospect it was found that the artificial walls for the CV were
placed too close to the periodic boundaries, so that atom jumps across
the boundary occurred, leading to heavy artifacts on the simulation
beyond 75,000 steps. A reasonably converged section of the FEP was
selected by finding an area where minima of equal free energy coincide
with O layers in the crystal. In this case, only two barriers could
be analyzed marked 1--2 and 2--3 in Fig.~\ref{fig:big-FEP}, yielding
a mean barrier of $1.09\pm0.13$~eV. This barrier is somewhat smaller
than in the MEP (cf. Tab.~\ref{tab:ME-Barriers}), which emphasizes the
necessity for free energy calculation even of this seemingly simple process.
 As in the 2x2x3 cell, all observed transitions
are C type transitions, reinforcing the observation that long range
migration of oxygen vacancies appears to be governed by this process.

\begin{figure}
\begin{centering}
\includegraphics[width=4cm]{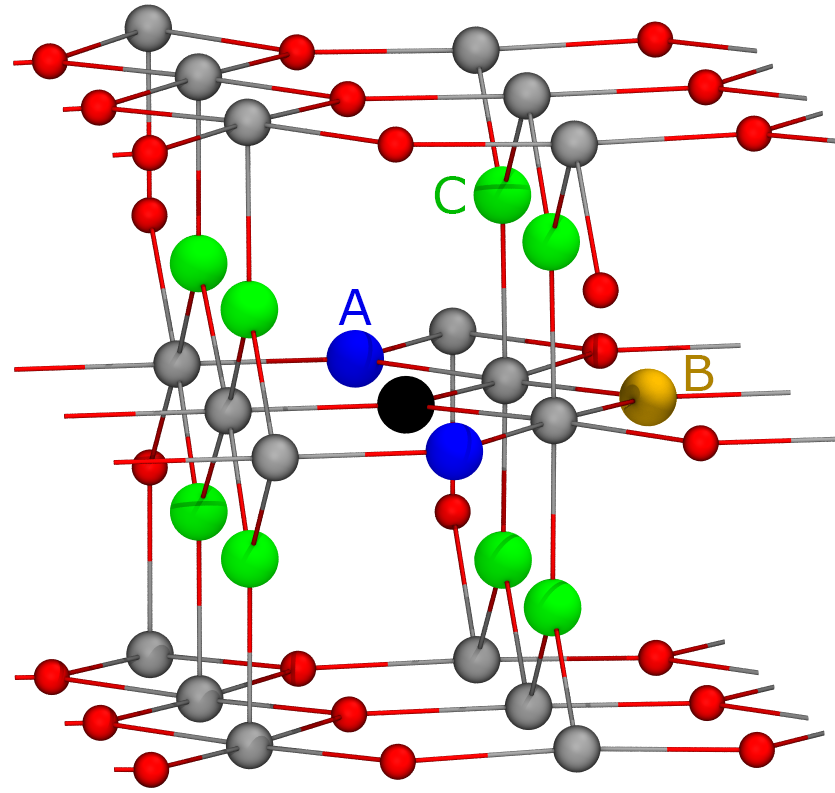}
\par\end{centering}

\caption{\label{fig:VO-transitions} (color on line) Possible transitions for
a V$_{\mathrm{O}}$ defect in rutile. Initial vacancy position is
marked in black.}

\end{figure}

For understanding the high preference for C jumps, it is necessary to
examine the possible transitions closer, this time centered on the
vacancy rather than a Ti atom. Fig.~\ref{fig:VO-transitions} shows
all end points of the three types of nearest neighbor jumps a vacancy
can undergo. The number of realizations for each event type varies
strongly, with the ratio of A:B:C being 2:1:8. The topology shows
that B jumps alone cannot lead to migration of the vacancy and A events
can only lead to strictly linear migration along the (001) axis. In
contrast, combinations of C jumps allow the vacancy to move in any
direction. A combination a B and C jumps can lead to accelerated movement
along the (110) and (1$\bar{1}$0) directions. From this follows,
that the rate-limiting step for vacancy diffusion in rutile is the
C-type jump, which is the only hopping event during which the vacancy
does not stay within the equatorial plane of any of its Ti neighbors.

In the current formulation, the PIVoT coordinate is limited to a single
vacancy, but extension to a multi-vacancy capable M-PIVoT formulation
for a known number of vacancies can be achieved in a straight-forward
manner by clustering the strongly weighted reference positions, e.g.
using the \emph{k-means} algorithm\cite{Hartigan1979}. Overcoming
the current limitation to a static, externally provided reference
lattice is equally straight forward, at least in the case of compounds.
Without loss of generality, regarding an AB compound, the expected
B positions can be inferred from the A sublattice. E.g., rutile is
composed of isosceles triangles with Ti vertices and edges of $\sim3.0$
and 2x$\sim3.6$~\AA~  in length. Reference positions of sufficient
quality can be calculated by taking the center of gravity of every
group of three Ti atoms no further apart than 3.6~\AA~  plus a commensurate
allowance for thermal fluctuation. Similar rules can be formulated
for any compound where the species of interest is at least 3-fold
coordinated.

Our simulation results demonstrate that meta dynamics are a useful
tool explore and understand diffusion mechanisms in complex materials.
In contrast to minimum energy path search approaches, where the start
and end configurations must be provided by an educated guess of the
researcher, and thus every possible process must be foreseen, no \emph{a-priori}
assumptions on the diffusion mechanism are imposed. The dangers of
this reliance on chemical intuition have been demonstrated, e.g. in
the case of surface adatom diffusion\cite{Liu1991334}. A collective
variable based (meta) dynamics approach avoids this pitfall that is
even deeper for vacancies, which are an emergent property of the distribution
if indistinguishable atoms.

Besides serving as a collective variable for free energy sampling,
the PIVoT coordinate also shows great potential for use as an event
detection method in parallel replica dynamics\cite{PhysRevB.57.R13985}
and related approaches. In contrast to the currently favored approach
of using differences in atom positions after steepest descent relaxation
of MD snapshots, the PIVoT coordinate is computationally so efficient
that it can be evaluated on the fly and events can be detected immediately,
reducing the computational cost while increasing the accuracy of event
time measurement.  Beyond tracking a vacancy outright, it is trivial to use
 the PIVoT coordinate
 as part of any other vector based variable to describe a system,
e.g. distances, angles, coordination numbers etc. The vacancy tracking 
is independent of the vacancy charge state and can be applied under any 
conditions as long as the interatomic potential remains valid. Combined with
Born-Oppenheimer dynamics, as shown here, the charge distribution is always
determined by the ionic arrangement. This allows to analyze correlations between
ion dynamics and electronic rearrangements.

Applied to the behavior of V$^{2+}_{\mathrm{O}}$ defects in rutile, we
demonstrated that great care must be taken in the choice of models
and computational details. Furthermore, our free energy surface samplings
show that not the lowest barrier pathway dominates the V$^{2+}_{\mathrm{O}}$
diffusion, but the C hop, which possesses the largest number of realizations
per O site and provides the largest connectivity between lattice sites.
While the good agreement between MEP and free energy paths in thermal
equilibrium is not very surprising in a crystalline solid, our results
demonstrate that free energy path sampling is in deed feasible in
the complex system of reduced titania. This opens the door to answering
much more challenging questions regarding the effect of external driving
forces such as electric fields or temperature gradients and their
relative impact on the memristive effect in different materials, which
is crucial for understanding and designing memristor devices\cite{Yang2013}.

The PIVoT coordinate was shown to be well suited to drive meta dynamics
simulation of oxygen vacancy diffusion in titania and to analyze the
resulting trajectories. We demonstrate that in combination with potentials
that offer a good compromise between efficiency, accuracy and transferability
such as SCC-DFTB, this vacancy coordinate allows examining the free
energy landscape of vacancy diffusion safe from the pitfalls of previously
existing mappings between atomic and vacancy motion. We show that
the diffusion of V$^{2+}_{\mathrm{O}}$ defects is dominated by out-of-plane
hopping events even though lower barrier jumps may occur.

The methods outlined in this work are applicable far beyond titania.
The PIVoT coordinate can be used to analyze and drive vacancy dynamics
in any solid, as long as the atomic arrangement does not undergo any
severe modification. This allows to tackle questions of vacancy dynamics
under external driving forces, such as electric fields or temperature
gradients. 

The authors thank Dr. Peter De\'ak of the BCCMS for discussions on the
simulation of V$_{\mathrm{O}}$ in TiO$_{2}$. We are grateful to
Mrs.~Svea gro\ss e Holthaus at the BCCMS and Professor Stephan Irle
at Nagoya University for advice about meta dynamics simulation. We
thank Dr. B\'alint Aradi at the BCCMS for his help in coupling the PLUMED
code to DFTB$^{+}$. We thank Prof. Stephen Elliott and Dr. Taehoon
Lee of the University of Cambridge for advice on void detection analysis.
All simulations were carried out using the DFTB$^{+}$ code\cite{dftbPlusURL},
meta dynamics simulation was performed by an extended version of the
PLUMED code\cite{Bonomi20091961} version 1.3. Model representations
were prepared using VMD\cite{HUMP96}. J.M.K is grateful for funding
from the DFG.

\bibliographystyle{apsrev}
\bibliography{PIVoT-ER-LV}

\begin{thebibliography}{41}
\expandafter\ifx\csname natexlab\endcsname\relax\def\natexlab#1{#1}\fi
\expandafter\ifx\csname bibnamefont\endcsname\relax
  \def\bibnamefont#1{#1}\fi
\expandafter\ifx\csname bibfnamefont\endcsname\relax
  \def\bibfnamefont#1{#1}\fi
\expandafter\ifx\csname citenamefont\endcsname\relax
  \def\citenamefont#1{#1}\fi
\expandafter\ifx\csname url\endcsname\relax
  \def\url#1{\texttt{#1}}\fi
\expandafter\ifx\csname urlprefix\endcsname\relax\def\urlprefix{URL }\fi
\providecommand{\bibinfo}[2]{#2}
\providecommand{\eprint}[2][]{\url{#2}}

\bibitem[{\citenamefont{Kwon et~al.}(2010)\citenamefont{Kwon, Kim, Jang, Jeon,
  Lee, Kim, Li, Park, Lee, Han et~al.}}]{Kwon-nn-5-148-2010}
\bibinfo{author}{\bibfnamefont{D.-H.} \bibnamefont{Kwon}},
  \bibinfo{author}{\bibfnamefont{K.~M.} \bibnamefont{Kim}},
  \bibinfo{author}{\bibfnamefont{J.~H.} \bibnamefont{Jang}},
  \bibinfo{author}{\bibfnamefont{J.~M.} \bibnamefont{Jeon}},
  \bibinfo{author}{\bibfnamefont{M.~H.} \bibnamefont{Lee}},
  \bibinfo{author}{\bibfnamefont{G.~H.} \bibnamefont{Kim}},
  \bibinfo{author}{\bibfnamefont{X.-S.} \bibnamefont{Li}},
  \bibinfo{author}{\bibfnamefont{G.-S.} \bibnamefont{Park}},
  \bibinfo{author}{\bibfnamefont{B.}~\bibnamefont{Lee}},
  \bibinfo{author}{\bibfnamefont{S.}~\bibnamefont{Han}}, \bibnamefont{et~al.},
  \bibinfo{journal}{nature nanotechnology} \textbf{\bibinfo{volume}{5}},
  \bibinfo{pages}{148} (\bibinfo{year}{2010}).

\bibitem[{\citenamefont{Park et~al.}(2011)\citenamefont{Park, Magyari-K\"ope,
  and Nishi}}]{Park-IEDL-32-197-2011}
\bibinfo{author}{\bibfnamefont{S.-G.} \bibnamefont{Park}},
  \bibinfo{author}{\bibfnamefont{B.}~\bibnamefont{Magyari-K\"ope}},
  \bibnamefont{and} \bibinfo{author}{\bibfnamefont{Y.}~\bibnamefont{Nishi}},
  \bibinfo{journal}{IEEE Electron Device Letters}
  \textbf{\bibinfo{volume}{32}}, \bibinfo{pages}{197} (\bibinfo{year}{2011}).

\bibitem[{\citenamefont{Strukov et~al.}(2008)\citenamefont{Strukov, Snider,
  Stewart, and Williams}}]{Strukov-N-453-80-2008}
\bibinfo{author}{\bibfnamefont{D.~B.} \bibnamefont{Strukov}},
  \bibinfo{author}{\bibfnamefont{G.~S.} \bibnamefont{Snider}},
  \bibinfo{author}{\bibfnamefont{D.~R.} \bibnamefont{Stewart}},
  \bibnamefont{and} \bibinfo{author}{\bibfnamefont{R.~S.}
  \bibnamefont{Williams}}, \bibinfo{journal}{Nature}
  \textbf{\bibinfo{volume}{453}}, \bibinfo{pages}{80} (\bibinfo{year}{2008}),
  \urlprefix\url{http://dx.doi.org/10.1038/nature06932}.

\bibitem[{\citenamefont{Strachan et~al.}(2009)\citenamefont{Strachan, Yang,
  M\"unstermann, Scholl, Medeiros-Ribeiro, Stewart, and
  Williams}}]{0957-4484-20-48-485701}
\bibinfo{author}{\bibfnamefont{J.~P.} \bibnamefont{Strachan}},
  \bibinfo{author}{\bibfnamefont{J.~J.} \bibnamefont{Yang}},
  \bibinfo{author}{\bibfnamefont{R.}~\bibnamefont{M\"unstermann}},
  \bibinfo{author}{\bibfnamefont{A.}~\bibnamefont{Scholl}},
  \bibinfo{author}{\bibfnamefont{G.}~\bibnamefont{Medeiros-Ribeiro}},
  \bibinfo{author}{\bibfnamefont{D.~R.} \bibnamefont{Stewart}},
  \bibnamefont{and} \bibinfo{author}{\bibfnamefont{R.~S.}
  \bibnamefont{Williams}}, \bibinfo{journal}{Nanotechnology}
  \textbf{\bibinfo{volume}{20}}, \bibinfo{pages}{485701}
  (\bibinfo{year}{2009}),
  \urlprefix\url{http://stacks.iop.org/0957-4484/20/i=48/a=485701}.

\bibitem[{\citenamefont{Waser et~al.}(2009)\citenamefont{Waser, Dittmann,
  Staikov, and Szot}}]{Waser-AM-21-2632-2009}
\bibinfo{author}{\bibfnamefont{R.}~\bibnamefont{Waser}},
  \bibinfo{author}{\bibfnamefont{R.}~\bibnamefont{Dittmann}},
  \bibinfo{author}{\bibfnamefont{G.}~\bibnamefont{Staikov}}, \bibnamefont{and}
  \bibinfo{author}{\bibfnamefont{K.}~\bibnamefont{Szot}},
  \bibinfo{journal}{Adv. Mater.} \textbf{\bibinfo{volume}{21}},
  \bibinfo{pages}{2632} (\bibinfo{year}{2009}),
  \urlprefix\url{http://dx.doi.org/10.1002/adma.200900375}.

\bibitem[{\citenamefont{Pickett et~al.}(2009)\citenamefont{Pickett, Strukov,
  Borghetti, Yang, Snider, Stewart, and Williams}}]{pickett:074508}
\bibinfo{author}{\bibfnamefont{M.~D.} \bibnamefont{Pickett}},
  \bibinfo{author}{\bibfnamefont{D.~B.} \bibnamefont{Strukov}},
  \bibinfo{author}{\bibfnamefont{J.~L.} \bibnamefont{Borghetti}},
  \bibinfo{author}{\bibfnamefont{J.~J.} \bibnamefont{Yang}},
  \bibinfo{author}{\bibfnamefont{G.~S.} \bibnamefont{Snider}},
  \bibinfo{author}{\bibfnamefont{D.~R.} \bibnamefont{Stewart}},
  \bibnamefont{and} \bibinfo{author}{\bibfnamefont{R.~S.}
  \bibnamefont{Williams}}, \bibinfo{journal}{J. Appl. Phys.}
  \textbf{\bibinfo{volume}{106}}, \bibinfo{eid}{074508}
  (pages~\bibinfo{numpages}{6}) (\bibinfo{year}{2009}),
  \urlprefix\url{http://link.aip.org/link/?JAP/106/074508/1}.

\bibitem[{\citenamefont{Bursill and Hyde}(1972)}]{Bursill-PSSC-7-177-1972}
\bibinfo{author}{\bibfnamefont{L.~A.} \bibnamefont{Bursill}} \bibnamefont{and}
  \bibinfo{author}{\bibfnamefont{B.~G.} \bibnamefont{Hyde}},
  \bibinfo{journal}{Prog. Sol. Stat. Chem.} \textbf{\bibinfo{volume}{7}},
  \bibinfo{pages}{177} (\bibinfo{year}{1972}).

\bibitem[{\citenamefont{Marezio et~al.}(1973)\citenamefont{Marezio, McWhan,
  Dernier, and Remeika}}]{Marezio1973213}
\bibinfo{author}{\bibfnamefont{M.}~\bibnamefont{Marezio}},
  \bibinfo{author}{\bibfnamefont{D.~B.} \bibnamefont{McWhan}},
  \bibinfo{author}{\bibfnamefont{P.~D.} \bibnamefont{Dernier}},
  \bibnamefont{and} \bibinfo{author}{\bibfnamefont{J.~P.}
  \bibnamefont{Remeika}}, \bibinfo{journal}{J. Solid State Chem.}
  \textbf{\bibinfo{volume}{6}}, \bibinfo{pages}{213 } (\bibinfo{year}{1973}),
  ISSN \bibinfo{issn}{0022-4596},
  \urlprefix\url{http://www.sciencedirect.com/science/article/pii/0022459673901849}.

\bibitem[{\citenamefont{Liborio and Harrison}(2008)}]{PhysRevB.77.104104}
\bibinfo{author}{\bibfnamefont{L.}~\bibnamefont{Liborio}} \bibnamefont{and}
  \bibinfo{author}{\bibfnamefont{N.}~\bibnamefont{Harrison}},
  \bibinfo{journal}{Phys. Rev. B} \textbf{\bibinfo{volume}{77}},
  \bibinfo{pages}{104104} (\bibinfo{year}{2008}).

\bibitem[{\citenamefont{Liborio et~al.}(2009)\citenamefont{Liborio, Mallia, and
  Harrison}}]{PhysRevB.79.245133}
\bibinfo{author}{\bibfnamefont{L.}~\bibnamefont{Liborio}},
  \bibinfo{author}{\bibfnamefont{G.}~\bibnamefont{Mallia}}, \bibnamefont{and}
  \bibinfo{author}{\bibfnamefont{N.}~\bibnamefont{Harrison}},
  \bibinfo{journal}{Phys. Rev. B} \textbf{\bibinfo{volume}{79}},
  \bibinfo{pages}{245133} (\bibinfo{year}{2009}).

\bibitem[{\citenamefont{Campbell and Peden}(2005)}]{Campbell29072005}
\bibinfo{author}{\bibfnamefont{C.~T.} \bibnamefont{Campbell}} \bibnamefont{and}
  \bibinfo{author}{\bibfnamefont{C.~H.~F.} \bibnamefont{Peden}},
  \bibinfo{journal}{Science} \textbf{\bibinfo{volume}{309}},
  \bibinfo{pages}{713} (\bibinfo{year}{2005}),
  \eprint{http://www.sciencemag.org/content/309/5735/713.full.pdf},
  \urlprefix\url{http://www.sciencemag.org/content/309/5735/713.short}.

\bibitem[{\citenamefont{Bikondoa et~al.}(2006/)\citenamefont{Bikondoa, Pang,
  Ithnin, Muryn, Onishi, and Thornton}}]{Bikondoa2006/}
\bibinfo{author}{\bibfnamefont{O.}~\bibnamefont{Bikondoa}},
  \bibinfo{author}{\bibfnamefont{C.~L.} \bibnamefont{Pang}},
  \bibinfo{author}{\bibfnamefont{R.}~\bibnamefont{Ithnin}},
  \bibinfo{author}{\bibfnamefont{C.~A.} \bibnamefont{Muryn}},
  \bibinfo{author}{\bibfnamefont{H.}~\bibnamefont{Onishi}}, \bibnamefont{and}
  \bibinfo{author}{\bibfnamefont{G.}~\bibnamefont{Thornton}},
  \bibinfo{journal}{Nat Mater} \textbf{\bibinfo{volume}{5}},
  \bibinfo{pages}{189} (\bibinfo{year}{2006/}),
  \urlprefix\url{http://dx.doi.org/10.1038/nmat1592}.

\bibitem[{\citenamefont{Gai-Boyes}(1992)}]{doi:10.1080/01614949208021918}
\bibinfo{author}{\bibfnamefont{P.~L.} \bibnamefont{Gai-Boyes}},
  \bibinfo{journal}{Catalysis Reviews} \textbf{\bibinfo{volume}{34}},
  \bibinfo{pages}{1} (\bibinfo{year}{1992}),
  \eprint{http://www.tandfonline.com/doi/pdf/10.1080/01614949208021918},
  \urlprefix\url{http://www.tandfonline.com/doi/abs/10.1080/01614949208021918}.

\bibitem[{\citenamefont{Wanbayor et~al.}(2012)\citenamefont{Wanbayor, De\'ak,
  Frauenheim, and Ruangpornvisuti}}]{Wanbayor201224}
\bibinfo{author}{\bibfnamefont{R.}~\bibnamefont{Wanbayor}},
  \bibinfo{author}{\bibfnamefont{P.}~\bibnamefont{De\'ak}},
  \bibinfo{author}{\bibfnamefont{T.}~\bibnamefont{Frauenheim}},
  \bibnamefont{and}
  \bibinfo{author}{\bibfnamefont{V.}~\bibnamefont{Ruangpornvisuti}},
  \bibinfo{journal}{Computational Materials Science}
  \textbf{\bibinfo{volume}{58}}, \bibinfo{pages}{24 } (\bibinfo{year}{2012}),
  ISSN \bibinfo{issn}{0927-0256},
  \urlprefix\url{http://www.sciencedirect.com/science/article/pii/S092702561200033X}.

\bibitem[{\citenamefont{Wanbayor et~al.}(2011)\citenamefont{Wanbayor, De\'{a}k,
  Frauenheim, and Ruangpornvisuti}}]{wanbayor:104701}
\bibinfo{author}{\bibfnamefont{R.}~\bibnamefont{Wanbayor}},
  \bibinfo{author}{\bibfnamefont{P.}~\bibnamefont{De\'{a}k}},
  \bibinfo{author}{\bibfnamefont{T.}~\bibnamefont{Frauenheim}},
  \bibnamefont{and}
  \bibinfo{author}{\bibfnamefont{V.}~\bibnamefont{Ruangpornvisuti}},
  \bibinfo{journal}{The Journal of Chemical Physics}
  \textbf{\bibinfo{volume}{134}}, \bibinfo{eid}{104701}
  (pages~\bibinfo{numpages}{6}) (\bibinfo{year}{2011}),
  \urlprefix\url{http://link.aip.org/link/?JCP/134/104701/1}.

\bibitem[{\citenamefont{Scheiber et~al.}(2012)\citenamefont{Scheiber, Fidler,
  Dulub, Schmid, Diebold, Hou, Aschauer, and Selloni}}]{PhysRevLett.109.136103}
\bibinfo{author}{\bibfnamefont{P.}~\bibnamefont{Scheiber}},
  \bibinfo{author}{\bibfnamefont{M.}~\bibnamefont{Fidler}},
  \bibinfo{author}{\bibfnamefont{O.}~\bibnamefont{Dulub}},
  \bibinfo{author}{\bibfnamefont{M.}~\bibnamefont{Schmid}},
  \bibinfo{author}{\bibfnamefont{U.}~\bibnamefont{Diebold}},
  \bibinfo{author}{\bibfnamefont{W.}~\bibnamefont{Hou}},
  \bibinfo{author}{\bibfnamefont{U.}~\bibnamefont{Aschauer}}, \bibnamefont{and}
  \bibinfo{author}{\bibfnamefont{A.}~\bibnamefont{Selloni}},
  \bibinfo{journal}{Phys. Rev. Lett.} \textbf{\bibinfo{volume}{109}},
  \bibinfo{pages}{136103} (\bibinfo{year}{2012}),
  \urlprefix\url{http://link.aps.org/doi/10.1103/PhysRevLett.109.136103}.

\bibitem[{\citenamefont{Dow et~al.}(1996)\citenamefont{Dow, Wang, and
  Huang}}]{Dow1996155}
\bibinfo{author}{\bibfnamefont{W.-P.} \bibnamefont{Dow}},
  \bibinfo{author}{\bibfnamefont{Y.-P.} \bibnamefont{Wang}}, \bibnamefont{and}
  \bibinfo{author}{\bibfnamefont{T.-J.} \bibnamefont{Huang}},
  \bibinfo{journal}{Journal of Catalysis} \textbf{\bibinfo{volume}{160}},
  \bibinfo{pages}{155 } (\bibinfo{year}{1996}), ISSN \bibinfo{issn}{0021-9517},
  \urlprefix\url{http://www.sciencedirect.com/science/article/pii/S0021951796901359}.

\bibitem[{\citenamefont{Laio and Parrinello}(2002)}]{Laio01102002}
\bibinfo{author}{\bibfnamefont{A.}~\bibnamefont{Laio}} \bibnamefont{and}
  \bibinfo{author}{\bibfnamefont{M.}~\bibnamefont{Parrinello}},
  \bibinfo{journal}{Proceedings of the National Academy of Sciences}
  \textbf{\bibinfo{volume}{99}}, \bibinfo{pages}{12562} (\bibinfo{year}{2002}),
  \eprint{http://www.pnas.org/content/99/20/12562.full.pdf+html},
  \urlprefix\url{http://www.pnas.org/content/99/20/12562.abstract}.

\bibitem[{\citenamefont{Gobbo et~al.}(2012)\citenamefont{Gobbo, Laio, Maleki,
  and Baroni}}]{PhysRevLett.109.150601}
\bibinfo{author}{\bibfnamefont{G.}~\bibnamefont{Gobbo}},
  \bibinfo{author}{\bibfnamefont{A.}~\bibnamefont{Laio}},
  \bibinfo{author}{\bibfnamefont{A.}~\bibnamefont{Maleki}}, \bibnamefont{and}
  \bibinfo{author}{\bibfnamefont{S.}~\bibnamefont{Baroni}},
  \bibinfo{journal}{Phys. Rev. Lett.} \textbf{\bibinfo{volume}{109}},
  \bibinfo{pages}{150601} (\bibinfo{year}{2012}),
  \urlprefix\url{http://link.aps.org/doi/10.1103/PhysRevLett.109.150601}.

\bibitem[{\citenamefont{D\'iaz~Leines and
  Ensing}(2012)}]{PhysRevLett.109.020601}
\bibinfo{author}{\bibfnamefont{G.}~\bibnamefont{D\'iaz~Leines}}
  \bibnamefont{and} \bibinfo{author}{\bibfnamefont{B.}~\bibnamefont{Ensing}},
  \bibinfo{journal}{Phys. Rev. Lett.} \textbf{\bibinfo{volume}{109}},
  \bibinfo{pages}{020601} (\bibinfo{year}{2012}),
  \urlprefix\url{http://link.aps.org/doi/10.1103/PhysRevLett.109.020601}.

\bibitem[{\citenamefont{Voter}(1998)}]{PhysRevB.57.R13985}
\bibinfo{author}{\bibfnamefont{A.~F.} \bibnamefont{Voter}},
  \bibinfo{journal}{Phys. Rev. B} \textbf{\bibinfo{volume}{57}},
  \bibinfo{pages}{R13985} (\bibinfo{year}{1998}),
  \urlprefix\url{http://link.aps.org/doi/10.1103/PhysRevB.57.R13985}.

\bibitem[{\citenamefont{Iddir et~al.}(2007)\citenamefont{Iddir,
  \"O\ifmmode~\breve{g}\else \u{g}\fi{}\"ut, Zapol, and
  Browning}}]{PhysRevB.75.073203}
\bibinfo{author}{\bibfnamefont{H.}~\bibnamefont{Iddir}},
  \bibinfo{author}{\bibfnamefont{S.}~\bibnamefont{\"O\ifmmode~\breve{g}\else
  \u{g}\fi{}\"ut}}, \bibinfo{author}{\bibfnamefont{P.}~\bibnamefont{Zapol}},
  \bibnamefont{and} \bibinfo{author}{\bibfnamefont{N.~D.}
  \bibnamefont{Browning}}, \bibinfo{journal}{Phys. Rev. B}
  \textbf{\bibinfo{volume}{75}}, \bibinfo{pages}{073203}
  (\bibinfo{year}{2007}).

\bibitem[{\citenamefont{Asaduzzaman and Kr\"uger}(2010)}]{Asaduzzaman2010}
\bibinfo{author}{\bibfnamefont{A.~M.} \bibnamefont{Asaduzzaman}}
  \bibnamefont{and} \bibinfo{author}{\bibfnamefont{P.}~\bibnamefont{Kr\"uger}},
  \bibinfo{journal}{J. Phys. Chem. C} \textbf{\bibinfo{volume}{114}},
  \bibinfo{pages}{19649} (\bibinfo{year}{2010}).

\bibitem[{\citenamefont{Knaup et~al.}(2012)\citenamefont{Knaup, Wehlau, and
  Frauenheim}}]{KnaupMRS2012}
\bibinfo{author}{\bibfnamefont{J.~M.} \bibnamefont{Knaup}},
  \bibinfo{author}{\bibfnamefont{M.}~\bibnamefont{Wehlau}}, \bibnamefont{and}
  \bibinfo{author}{\bibfnamefont{T.}~\bibnamefont{Frauenheim}}, in
  \emph{\bibinfo{booktitle}{MRS Proceedings}} (\bibinfo{year}{2012}), vol.
  \bibinfo{volume}{1430}, pp. \bibinfo{pages}{1430mrss12--1430--e08--10}.

\bibitem[{\citenamefont{Jug et~al.}(2005)\citenamefont{Jug, Nair, and
  Bredow}}]{B502507A}
\bibinfo{author}{\bibfnamefont{K.}~\bibnamefont{Jug}},
  \bibinfo{author}{\bibfnamefont{N.~N.} \bibnamefont{Nair}}, \bibnamefont{and}
  \bibinfo{author}{\bibfnamefont{T.}~\bibnamefont{Bredow}},
  \bibinfo{journal}{Phys. Chem. Chem. Phys.} \textbf{\bibinfo{volume}{7}},
  \bibinfo{pages}{2616} (\bibinfo{year}{2005}),
  \urlprefix\url{http://dx.doi.org/10.1039/B502507A}.

\bibitem[{\citenamefont{Akola and Jones}(2007)}]{prb-76-235201-2007}
\bibinfo{author}{\bibfnamefont{J.}~\bibnamefont{Akola}} \bibnamefont{and}
  \bibinfo{author}{\bibfnamefont{R.}~\bibnamefont{Jones}},
  \bibinfo{journal}{phys. rev. B} \textbf{\bibinfo{volume}{76}},
  \bibinfo{pages}{235201} (\bibinfo{year}{2007}).

\bibitem[{\citenamefont{Lee and Elliott}(2011)}]{PhysRevB.84.094124}
\bibinfo{author}{\bibfnamefont{T.~H.} \bibnamefont{Lee}} \bibnamefont{and}
  \bibinfo{author}{\bibfnamefont{S.~R.} \bibnamefont{Elliott}},
  \bibinfo{journal}{Phys. Rev. B} \textbf{\bibinfo{volume}{84}},
  \bibinfo{pages}{094124} (\bibinfo{year}{2011}),
  \urlprefix\url{http://link.aps.org/doi/10.1103/PhysRevB.84.094124}.

\bibitem[{Note1()}]{Note1}
Note1, \bibinfo{note}{for the sake of simplicity and without loss of generality
  we assume for a moment that the transition barrier is symmetrical between the
  adjacent minima.}

\bibitem[{\citenamefont{Frauenheim et~al.}(2000)\citenamefont{Frauenheim,
  Seifert, Elstner, Hajnal, Jungnickel, Porezag, Suhai, and
  Scholz}}]{ThFRAUENHEIM-GS-ME-ZH-GJ-DP-SS-RS-PSSb-217-41}
\bibinfo{author}{\bibfnamefont{T.}~\bibnamefont{Frauenheim}},
  \bibinfo{author}{\bibfnamefont{G.}~\bibnamefont{Seifert}},
  \bibinfo{author}{\bibfnamefont{M.}~\bibnamefont{Elstner}},
  \bibinfo{author}{\bibfnamefont{Z.}~\bibnamefont{Hajnal}},
  \bibinfo{author}{\bibfnamefont{G.}~\bibnamefont{Jungnickel}},
  \bibinfo{author}{\bibfnamefont{D.}~\bibnamefont{Porezag}},
  \bibinfo{author}{\bibfnamefont{S.}~\bibnamefont{Suhai}}, \bibnamefont{and}
  \bibinfo{author}{\bibfnamefont{R.}~\bibnamefont{Scholz}},
  \bibinfo{journal}{phys. stat. sol. (b)} \textbf{\bibinfo{volume}{217}},
  \bibinfo{pages}{41} (\bibinfo{year}{2000}).

\bibitem[{\citenamefont{Aradi et~al.}(2007)\citenamefont{Aradi, Hourahine, and
  Frauenheim}}]{JPCA-111-5678-2007}
\bibinfo{author}{\bibfnamefont{B.}~\bibnamefont{Aradi}},
  \bibinfo{author}{\bibfnamefont{B.}~\bibnamefont{Hourahine}},
  \bibnamefont{and}
  \bibinfo{author}{\bibfnamefont{T.}~\bibnamefont{Frauenheim}},
  \bibinfo{journal}{J. Phys. Chem. A} \textbf{\bibinfo{volume}{111}},
  \bibinfo{pages}{5678} (\bibinfo{year}{2007}).

\bibitem[{\citenamefont{Dolgonos et~al.}(2010)\citenamefont{Dolgonos, Aradi,
  Moreira, and Frauenheim}}]{doi:10.1021/ct900422c}
\bibinfo{author}{\bibfnamefont{G.}~\bibnamefont{Dolgonos}},
  \bibinfo{author}{\bibfnamefont{B.}~\bibnamefont{Aradi}},
  \bibinfo{author}{\bibfnamefont{N.~H.} \bibnamefont{Moreira}},
  \bibnamefont{and}
  \bibinfo{author}{\bibfnamefont{T.}~\bibnamefont{Frauenheim}},
  \bibinfo{journal}{Journal of Chemical Theory and Computation}
  \textbf{\bibinfo{volume}{6}}, \bibinfo{pages}{266} (\bibinfo{year}{2010}),
  \eprint{http://pubs.acs.org/doi/pdf/10.1021/ct900422c},
  \urlprefix\url{http://pubs.acs.org/doi/abs/10.1021/ct900422c}.

\bibitem[{\citenamefont{Monkhorst and Pack}(1976)}]{PhysRevB.13.5188}
\bibinfo{author}{\bibfnamefont{H.~J.} \bibnamefont{Monkhorst}}
  \bibnamefont{and} \bibinfo{author}{\bibfnamefont{J.~D.} \bibnamefont{Pack}},
  \bibinfo{journal}{Phys. Rev. B} \textbf{\bibinfo{volume}{13}},
  \bibinfo{pages}{5188} (\bibinfo{year}{1976}),
  \urlprefix\url{http://link.aps.org/doi/10.1103/PhysRevB.13.5188}.

\bibitem[{\citenamefont{De\'ak et~al.}(2012)\citenamefont{De\'ak, Aradi, and
  Frauenheim}}]{PhysRevB.86.195206}
\bibinfo{author}{\bibfnamefont{P.}~\bibnamefont{De\'ak}},
  \bibinfo{author}{\bibfnamefont{B.}~\bibnamefont{Aradi}}, \bibnamefont{and}
  \bibinfo{author}{\bibfnamefont{T.}~\bibnamefont{Frauenheim}},
  \bibinfo{journal}{Phys. Rev. B} \textbf{\bibinfo{volume}{86}},
  \bibinfo{pages}{195206} (\bibinfo{year}{2012}),
  \urlprefix\url{http://link.aps.org/doi/10.1103/PhysRevB.86.195206}.

\bibitem[{\citenamefont{Nos\'{e}}(1984)}]{Nose1984}
\bibinfo{author}{\bibfnamefont{S.}~\bibnamefont{Nos\'{e}}},
  \bibinfo{journal}{J. Chem. Phys.} \textbf{\bibinfo{volume}{81}},
  \bibinfo{pages}{511} (\bibinfo{year}{1984}),
  \urlprefix\url{http://link.aip.org/link/?JCP/81/511/1}.

\bibitem[{\citenamefont{Hoover}(1985)}]{PhysRevA.31.1695}
\bibinfo{author}{\bibfnamefont{W.~G.} \bibnamefont{Hoover}},
  \bibinfo{journal}{Phys. Rev. A} \textbf{\bibinfo{volume}{31}},
  \bibinfo{pages}{1695} (\bibinfo{year}{1985}),
  \urlprefix\url{http://link.aps.org/doi/10.1103/PhysRevA.31.1695}.

\bibitem[{\citenamefont{Hartigan and Wong}(1979)}]{Hartigan1979}
\bibinfo{author}{\bibfnamefont{J.~A.} \bibnamefont{Hartigan}} \bibnamefont{and}
  \bibinfo{author}{\bibfnamefont{M.~A.} \bibnamefont{Wong}},
  \bibinfo{journal}{Journal of the Royal Statistical Society. Series C (Applied
  Statistics)} \textbf{\bibinfo{volume}{28}}, \bibinfo{pages}{pp. 100}
  (\bibinfo{year}{1979}), ISSN \bibinfo{issn}{00359254},
  \urlprefix\url{http://www.jstor.org/stable/2346830}.

\bibitem[{\citenamefont{Liu et~al.}(1991)\citenamefont{Liu, Cohen, Adams, and
  Voter}}]{Liu1991334}
\bibinfo{author}{\bibfnamefont{C.}~\bibnamefont{Liu}},
  \bibinfo{author}{\bibfnamefont{J.}~\bibnamefont{Cohen}},
  \bibinfo{author}{\bibfnamefont{J.}~\bibnamefont{Adams}}, \bibnamefont{and}
  \bibinfo{author}{\bibfnamefont{A.}~\bibnamefont{Voter}},
  \bibinfo{journal}{Surface Science} \textbf{\bibinfo{volume}{253}},
  \bibinfo{pages}{334 } (\bibinfo{year}{1991}), ISSN \bibinfo{issn}{0039-6028},
  \urlprefix\url{http://www.sciencedirect.com/science/article/pii/003960289190604Q}.

\bibitem[{\citenamefont{Yang et~al.}(2013)\citenamefont{Yang, Strukov, and
  Stewart}}]{Yang2013}
\bibinfo{author}{\bibfnamefont{J.~J.} \bibnamefont{Yang}},
  \bibinfo{author}{\bibfnamefont{D.~B.} \bibnamefont{Strukov}},
  \bibnamefont{and} \bibinfo{author}{\bibfnamefont{D.~R.}
  \bibnamefont{Stewart}}, \bibinfo{journal}{Nat Nano}
  \textbf{\bibinfo{volume}{8}}, \bibinfo{pages}{13} (\bibinfo{year}{2013}),
  \urlprefix\url{http://dx.doi.org/10.1038/nnano.2012.240}.

\bibitem[{dft()}]{dftbPlusURL}
\emph{\bibinfo{title}{{DFTB}$^{+}$}},
  \urlprefix\url{http://www.dftb-plus.info/}.

\bibitem[{\citenamefont{Bonomi et~al.}(2009)\citenamefont{Bonomi, Branduardi,
  Bussi, Camilloni, Provasi, Raiteri, Donadio, Marinelli, Pietrucci, Broglia
  et~al.}}]{Bonomi20091961}
\bibinfo{author}{\bibfnamefont{M.}~\bibnamefont{Bonomi}},
  \bibinfo{author}{\bibfnamefont{D.}~\bibnamefont{Branduardi}},
  \bibinfo{author}{\bibfnamefont{G.}~\bibnamefont{Bussi}},
  \bibinfo{author}{\bibfnamefont{C.}~\bibnamefont{Camilloni}},
  \bibinfo{author}{\bibfnamefont{D.}~\bibnamefont{Provasi}},
  \bibinfo{author}{\bibfnamefont{P.}~\bibnamefont{Raiteri}},
  \bibinfo{author}{\bibfnamefont{D.}~\bibnamefont{Donadio}},
  \bibinfo{author}{\bibfnamefont{F.}~\bibnamefont{Marinelli}},
  \bibinfo{author}{\bibfnamefont{F.}~\bibnamefont{Pietrucci}},
  \bibinfo{author}{\bibfnamefont{R.~A.} \bibnamefont{Broglia}},
  \bibnamefont{et~al.}, \bibinfo{journal}{Computer Physics Communications}
  \textbf{\bibinfo{volume}{180}}, \bibinfo{pages}{1961 }
  (\bibinfo{year}{2009}), ISSN \bibinfo{issn}{0010-4655},
  \urlprefix\url{http://www.sciencedirect.com/science/article/pii/S001046550900157X}.

\bibitem[{\citenamefont{Humphrey et~al.}(1996)\citenamefont{Humphrey, Dalke,
  and Schulten}}]{HUMP96}
\bibinfo{author}{\bibfnamefont{W.}~\bibnamefont{Humphrey}},
  \bibinfo{author}{\bibfnamefont{A.}~\bibnamefont{Dalke}}, \bibnamefont{and}
  \bibinfo{author}{\bibfnamefont{K.}~\bibnamefont{Schulten}},
  \bibinfo{journal}{Journal of Molecular Graphics}
  \textbf{\bibinfo{volume}{14}}, \bibinfo{pages}{33} (\bibinfo{year}{1996}),
  \urlprefix\url{http://www.ks.uiuc.edu/Research/vmd/}.

\end{thebibliography}

\end{document}